\documentclass[11pt]{article}
\usepackage[T1]{fontenc}
\usepackage[utf8]{inputenc}
\usepackage{times}
\usepackage[letterpaper,left=1.5in,right=1in,top=1in,bottom=1in]{geometry}
\usepackage{graphicx,booktabs,caption,titlesec,microtype,adjustbox}
\usepackage{amssymb}
\usepackage{array}
\newcolumntype{L}[1]{>{\raggedright\arraybackslash}p{#1}}
\usepackage[authoryear,round]{natbib}
\usepackage[hidelinks,breaklinks]{hyperref}
\usepackage{changepage}
\usepackage{pdfpages}
\usepackage{float}
\DeclareUnicodeCharacter{2713}{\checkmark}

\providecommand{\figalttext}[2][]{}
\providecommand{\botrule}{\bottomrule}

\titleformat{\section}{\normalsize\scshape}{\thesection}{1em}{\scshape}
\titleformat{\subsection}{\normalsize\scshape}{\thesubsection}{1em}{\scshape}
\titlespacing*{\section}{0pt}{18pt}{8pt}
\titlespacing*{\subsection}{0pt}{12pt}{6pt}

\begin{document}

\begin{center}
{\large\scshape MechGate: combining mechanism-derived measures of permeation
and efflux to screen small molecule drugs targeting \textit{Escherichia coli}}
\end{center}

\vspace{6pt}
Lang Qin$^{1}$, Jagath C Rajapakse$^{1,\ast}$

{\footnotesize $^{1}$College of Computing and Data Science, Nanyang Technological University, Singapore}

{\footnotesize $^{\ast}$Corresponding author: \texttt{asJagath@ntu.edu.sg}}

\vspace{12pt}

\begin{center}\scshape Abstract\end{center}
\begin{adjustwidth}{0.4in}{0.4in}
\small Most antibiotic molecules must cross the cell membrane and escape efflux pumps before hitting their targets. For de novo antibiotic design, very few of the in-silico-generated drug molecules can successfully pass these two steps. Existing evaluations of generated molecules do not separate these steps. Therefore, we introduce MechGate, a mechanism-resolved screening tool evaluating small drug molecules against \textit{E. coli}, measuring permeation, efflux and antibacterial activity. After SMILES input, five measures are reported side by side: two for permeation via porins, one for permeation via the phospholipid membrane, one for efflux and one for activity. Each measure is read on its own, so a molecule that fails can be traced to the step that stops it. All features are taken from papers that state their mechanism rather than selected by importance ranking. On an external set of marketed antibacterials the measures give a precision of 0.789 to 0.873 against baselines near 0.44, and combining them the way the mechanism predicts raises precision to 0.92. We also find that the four public source libraries are separable from one another using physicochemical features alone; every measure here therefore draws both classes from a single library. All code is open-source and requires no commercial software. It is freely available at https://github.com/qinlang0008-blip/mechgate.
\end{adjustwidth}

\vspace{6pt}
\begin{adjustwidth}{0.4in}{0.4in}
\end{adjustwidth}

\vspace{12pt}
\begin{adjustwidth}{0.4in}{0.4in}
\small\textbf{Key Points}
\begin{itemize}
\item MechGate reports five mechanism-derived measures of permeation, efflux and whole-cell activity side by side for small molecules in \textit{E. coli}. No total score is formed and the molecules are not ranked.
\item Every feature is taken from a paper that states its mechanism, and mechanism also decides which measures are built and how they are combined.
\item On an external set of marketed antibacterials the five measures reach a precision of 0.789 to 0.873 against baselines near 0.44. Combining them as the mechanism predicts raises precision to 0.92.
\item Precision and recall are reported on a set drawn from a source other than the training data, which to our knowledge is uncommon for models of Gram-negative permeation and efflux.
\item Four public source libraries are separable from one another by physicochemical features alone, so validation inside a single library cannot show that a model has learnt chemistry rather than provenance.
\end{itemize}
\end{adjustwidth}

\vspace{12pt}
\section{Introduction}

Bacterial resistance is now a global threat, causing over a million deaths worldwide directly in 2021 alone \citep{gbd2024}. No clinically important new class has emerged since the quinolones \citep{elzahed2021}, and high-throughput screening of more than 8 million compounds has not produced a single marketed drug for Gram-negative infection \citep{prajapati2021}. Addressing these issues, de novo antibiotic design has drawn increasing attention. In recent years, machine learning has become the main approach in drug discovery, and in antibiotic design as well \citep{chenl2023,shen2024,kollen2025,krishnan2025}. Nevertheless, many of the molecules generated by these models in silico are not effective in practice. This makes downstream selection, or molecule evaluation, significant.

Tools and benchmarks for evaluation have developed for evaluation of molecules and generative models. Benchmarks score the molecules a model produces for validity, uniqueness, novelty and distribution match \citep{brownn2019,polykovskiy2020,thomas2024,chenl2023}, and later ones added docking against a target \citep{garciaortegon2022,cieplinski2023} and ADME and safety constraints \citep{jones2025}; the target, however, none of these scores the process whether a molecule can enter and stay in a cell for killing it. A standardised evaluation set for antibacterials exists for antimicrobial peptides \citep{lavertu2026}, instead of small molecules. When it comes to tools, eNTRyway predicts accumulation in \textit{E. coli} from structure \citep{richtermf2017,parker2020}, and a gradient-boosted model predicts retention from a tolC-paired LC/MS screen \citep{legoff2025}, both of which are permeation net of efflux; adding molecular-dynamics features improves accumulation prediction but requires a simulation per molecule \citep{milenkovic2024}. Only passage through the porins --- the water-filled channels of the outer membrane --- has been scored on its own \citep{acosta2018}. As a result, entry is often left out of screening altogether: in one published antibiotic screening pipeline, none of the five cascaded criteria concerns it \citep{krishnan2025}.

Before killing a bacterial cell, an antibiotic molecule must first pass through the cell membrane. This process is called permeation. There are three routes for permeation: via a porin, an aqueous channel (the porin route); across the phospholipid membrane itself (the lipid route), which is the only route open to lipophilic molecules but not restricted to them; and via self-promoted uptake for aminoglycosides and lipopeptides \citep{hancock1991}. Thus, if the physicochemical properties of a molecule meet the requirements of any one of the three routes, the molecule is likely to permeate. Each route, however, imposes different requirements \citep{richtermf2017,acosta2018,nikaido2003}. After permeation, some molecules are then pumped out of the cell again. This process is called efflux. In order to remain inside the membrane, a molecule must avoid being pumped out. Therefore, the lower its efflux rate, the higher its retention \citep{cooper2018,elzahed2021,legoff2025}. These steps are shown in Figure 1.

\begin{figure}[!t]
\centering
\includegraphics[width=\textwidth]{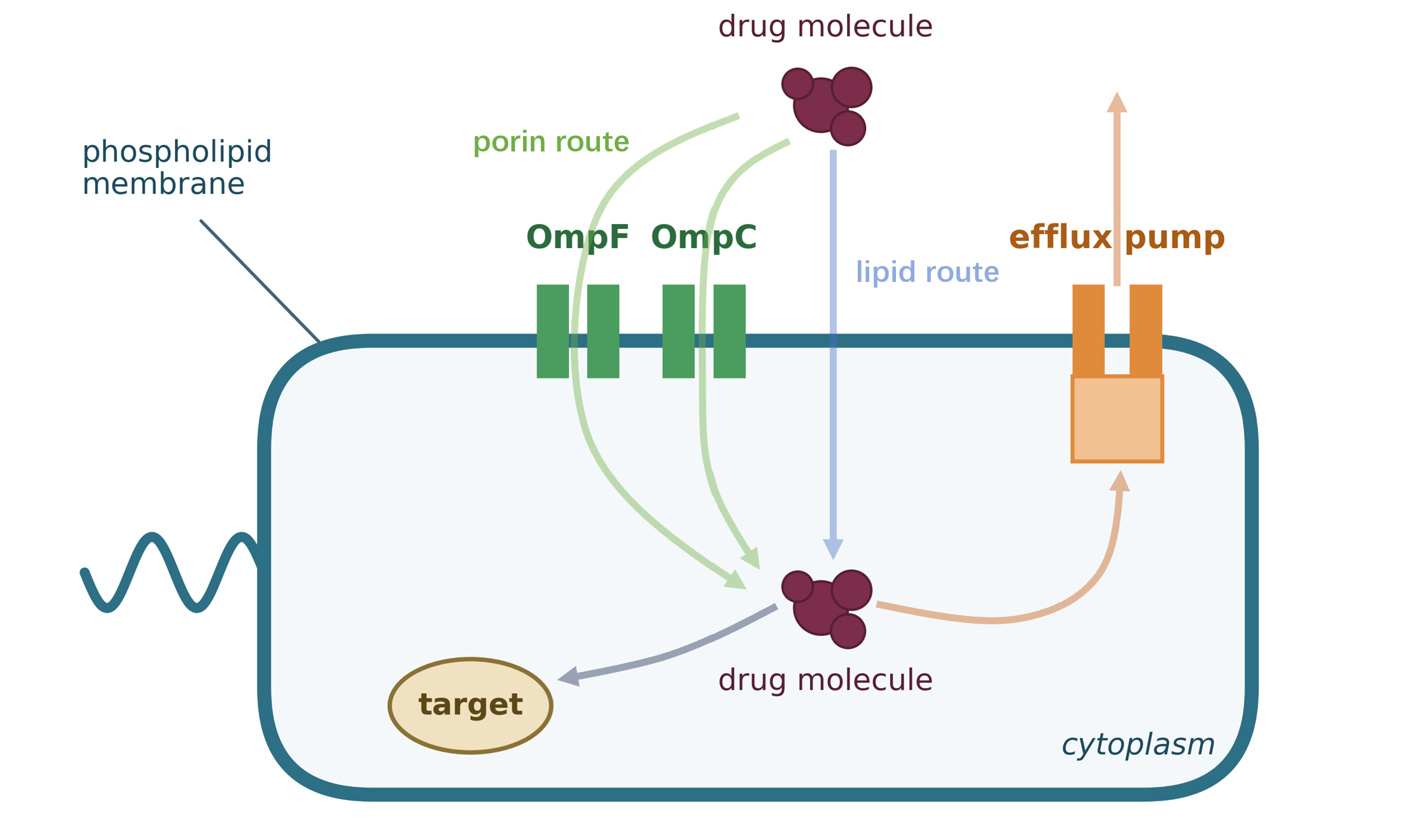}
\caption{How an antibiotic molecule reaches its target in \textit{E. coli}. A molecule crosses the membrane through the OmpF porin, through the OmpC porin, or across the membrane itself. Once inside, it may be pumped back out by an efflux pump, or it may reach its target. Green arrows mark the porin route, blue the lipid route, and orange the efflux process.}\label{fig1}
\figalttext[Routes into a Gram-negative cell]{Schematic of a Gram-negative cell. A drug molecule reaches the cytoplasm through the OmpF porin, through the OmpC porin, or across the phospholipid membrane. An efflux pump returns it to the outside, and a further arrow leads to the target.}
{\footnotesize Alt text: Schematic of a Gram-negative cell. A drug molecule reaches the cytoplasm through the OmpF porin, through the OmpC porin, or across the phospholipid membrane. An efflux pump returns it to the outside, and a further arrow leads to the target.}
\end{figure}

Alt text: Schematic of a Gram-negative cell. A drug molecule reaches the cytoplasm through the OmpF porin, through the OmpC porin, or across the phospholipid membrane. An efflux pump returns it to the outside, and a further arrow leads to the target.

Only after these two processes can an antibiotic molecule reach its target. Whole-cell readouts such as the MIC, however, show the joint outcome alone and cannot be attributed to any single step \citep{leus2023,elzahed2021}. So the effectiveness of a molecule depends largely on its permeation rate and its efflux rate. Existing work has followed three broad lines. The first writes down readable rules or explicit physical terms from mechanism, but the training sets are small \citep{acosta2018,richtermf2017}. The second builds statistical models on large screens, but selects features by importance ranking rather than by mechanism \citep{elzahed2021,legoff2025,cooper2018,gurvic2022,gurvic2024}. The third separates the endpoints, but either performs no external validation or builds no model on the compound side \citep{leus2023,teelucksingh2022,masi2022,krishnamoorthy2017,sugawara2016,scherer2025}. Two further studies measure something else again: genome-wide deletion libraries report growth fitness rather than permeability \citep{nichols2011}, and screening in efflux-deficient strains is very small and carried out in Pseudomonas aeruginosa \citep{urdanetapaez2023}. However, all of these works address only one of the two processes. El Zahed has noted that the respective contributions of outer-membrane permeability and of efflux remain unclear \citep{elzahed2021}. And none of them tested its models on external sets, which makes overfitting difficult to identify. 

This situation may be caused by data limitation. The permeation labels available across the whole field amount to little: 45 molecules with measured porin permeability coefficients, 84 from liposome experiments, and about 672 rows released in total by four antibiotic generative models, largely the same molecules. BindingDB holds 3.2 million target-binding records but no MIC (BindingDB), and antibacterial high-throughput screens have a hit rate of only 1--2\% and are mostly not public \citep{onawole2026}. The endpoint of each dataset is also different. Pooling several source libraries into one training set has precedents in this field \citep{gurvic2022,milenkovic2024,boi2025}. We have built a set of classification models to identify which dataset a molecule comes from, finding that all four are separable from one another using physicochemical features alone, with no permeation label. If a training set is built with the positive data from one source library and the negative data from another, we are not able to tell whether the model has learnt the label, or just the difference between the two original source libraries. For these reasons, the existing source libraries cannot be aligned into a single training set. Such a data limitation makes it nearly impossible to train, or to adjust, a generative model. So the most feasible way in drug design in silico is to select out the molecules that seem effective, using a dedicated screening tool.

In this work, we introduce MechGate, a mechanism-resolved screening tool evaluating small drug molecules against E. coli, measuring permeation, efflux and antibacterial activity, which may significantly reduce the in-vitro selection in the wet lab. After the user input the SMILES of a molecule, or a set of molecules, five measures are reported side by side: two for permeation via porins, one for permeation via the phospholipid membrane, one for efflux and one for activity. Each measure is read on its own, so a molecule that fails can be traced to the step that stops it. Our main contributions can be summarized as follows:
1, We have built MechGate, the first mechanism-derived model to our knowledge, that scores permeation and efflux as one integrated tool for small molecules in \textit{E. coli}. It reports whole-cell activity together with the separate steps that lead to it
2, We create a method for evaluating molecules from mechanism: mechanism decides measure choice, feature choice, and the way to combine the measures. Combining the measures the way the mechanism predicts raises precision on external data
3, This is the first model of Gram-negative permeation and efflux, to our knowledge, whose precision and recall are reported on a set drawn from a source other than its own training data.
4, Four public source libraries can be told apart from one another by physicochemical features alone, so a model validated within one library cannot be shown to have learnt chemistry rather than provenance.

\section{Materials and methods}

\subsection{Overview of MechGate}

The input is a single column of SMILES strings; SMILES is the only input required. The InChIKey in the output is computed from the SMILES, and the 3D conformer needed by the lipid route measure is generated from it as well, so neither has to be supplied by the user. The output is a 20-column table with one row per molecule.

A generative model first produces candidate molecules. Once these have passed the general benchmarks that check whether they look like drugs, this assessment can be applied. It is antibiotic-dedicated, and tests only whether a candidate is likely to be effective in permeation, efflux and activity.

The tool comprises three tracks and five measures, and the tracks follow the passage described above. Of the two membrane layers in a Gram-negative cell, only the outer membrane is modelled here, since existing measured data concern mainly that layer \citep{nikaido2003}. Of the three permeation routes, only the porin route and the lipid route are given measures; self-promoted uptake is used by two classes of molecule only and has too little published data to model, so those molecules are flagged instead. Permeation and efflux form the two main tracks. Two tracks are not enough, because they cover only what happens before a molecule reaches its target. A third track, the activity track, was therefore added.

The five measures are named the OmpF measure, the multi-porin measure, the lipid route measure, the efflux measure and the activity measure.

The tool therefore has three tracks and five measures. Three measures belong to the permeation track; the efflux and activity tracks have one each. Of the three permeation measures, two concern porins, which carry hydrophilic molecules, and the third concerns passage across the membrane itself, applying to lipophilic and hydrophilic molecules alike. Figure 2 shows how the five measures are read together.

\begin{figure}[!t]
\centering
\includegraphics[width=\textwidth]{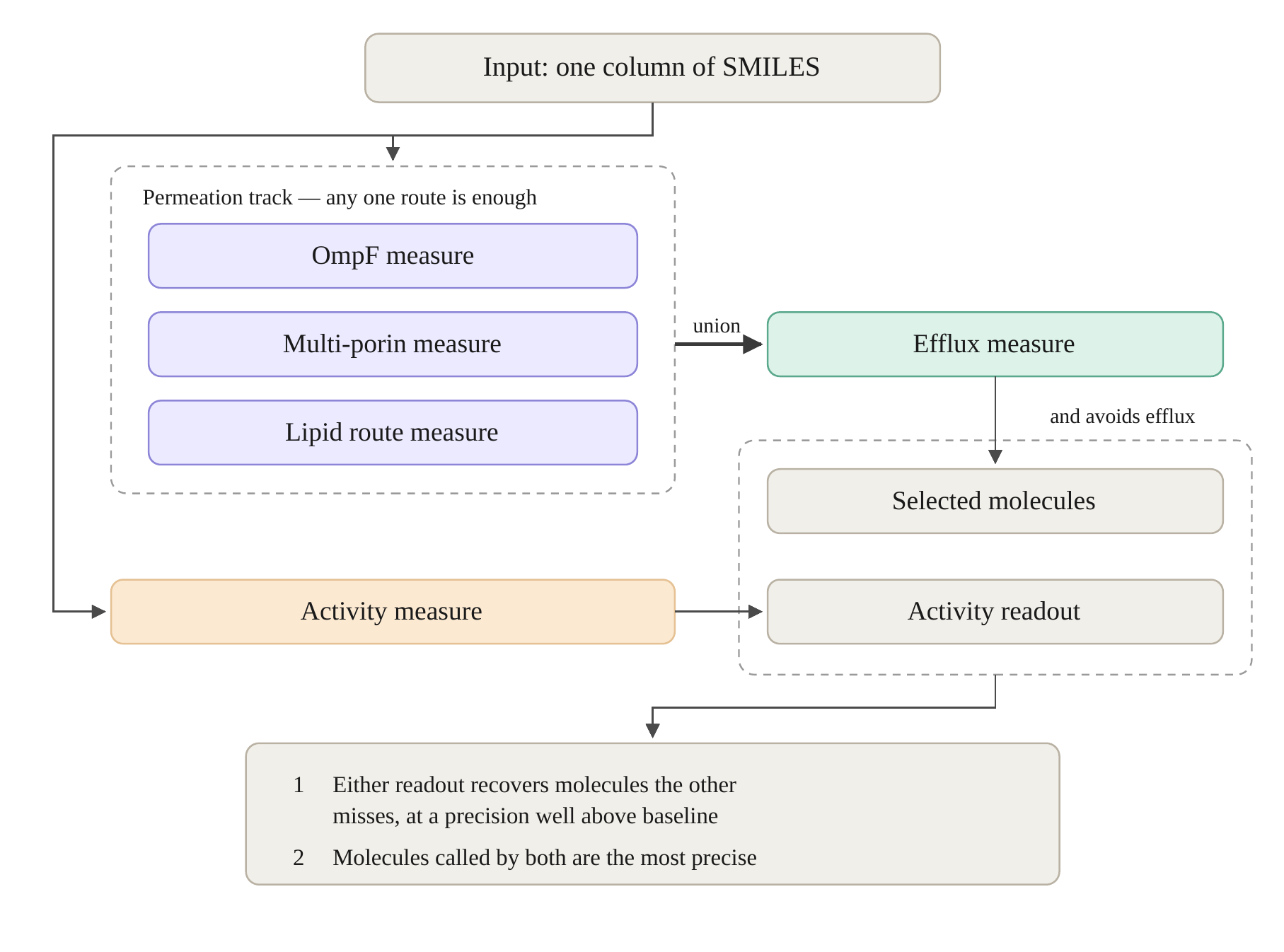}
\caption{The five measures and what each of them reports. The three permeation measures are combined by union, and a molecule is kept only if it also avoids efflux; the activity measure is reported separately. Statements 1 and 2 are read on the 442 molecules left after removing the training molecules of all five measures, where the baseline is 0.396: precision is 0.720 for the mechanism side alone (n = 25), 0.758 for the activity measure alone (n = 62) and 0.981 for the molecules called by both (n = 52). They hold where the activity measure is taken at the 0.95 percentile or tighter, and there at all seven efflux reference points (Note S11).}\label{fig2}
\figalttext[Flow chart of the five measures]{Flow chart. One column of SMILES enters a permeation track holding the OmpF, multi-porin and lipid route measures, which are combined by union and then intersected with the efflux measure to give selected molecules. The activity measure runs alongside and gives its own readout.}
{\footnotesize Alt text: Flow chart. One column of SMILES enters a permeation track holding the OmpF, multi-porin and lipid route measures, which are combined by union and then intersected with the efflux measure to give selected molecules. The activity measure runs alongside and gives its own readout.}
\end{figure}

Alt text: Flow chart. One column of SMILES enters a permeation track holding the OmpF, multi-porin and lipid route measures, which are combined by union and then intersected with the efflux measure to give selected molecules. The activity measure runs alongside and gives its own readout.

\subsection{Features and algorithms}

Each of the nine features is taken from a paper that states its mechanism, and they fall into four groups by the passage their source papers assign them to. One is assigned to the permeation side as a whole: molecular weight \citep{krishnamoorthy2017}. Two are assigned to the porin route: net charge at pH 7.4 and number of primary amines on sp3 carbon \citep{acosta2018,acosta2021,richtermf2017}. Two are assigned to the outer-membrane barrier: logP at pH 7.4 and fraction of rotatable bonds \citep{cooper2018}. Two are assigned to efflux: zwitterionic group flag and zwitterionic charge flag \citep{browndg2014}. The last two are transverse dipole variation across conformers and minimum total dipole moment across conformers. Both come from \citet{acosta2018}, and that paper states no passage for them. For the first of the two, the quantity comes from that paper, but the choice of a variation statistic over a mean follows \citet{boi2025}. The direction of minimum total dipole moment across conformers comes from the same paper, and the feature also appears in two independent groups \citep{ferreira2019,richterr2020}. An earlier version used 16 features. Seven were removed on four grounds, following recursive feature elimination \citep{guyon2002}, and the grounds for each are given in Note S1. All features were computed with RDKit \citep{landrum2023}, and no commercial software was used. Note S1 also gives the exact definition of each feature and the name it carries in the tool output.

This work uses existing algorithms and datasets to address the problem. The five measures do not share a single algorithm. Three algorithms were tried: XGBoost, Logistic Regression and Random Forest. All are implemented in scikit-learn \citep{pedregosa2011}. The OmpF measure is not trained at all, and the other four are. The algorithm used by each measure is given in that measure's own section below.

\subsection{OmpF measure}

The two porin measures differ in scope. One applies to OmpF alone, the porin most studied in \textit{E. coli} and its main channel for antibiotic uptake \citep{nikaido2003,acosta2018}. 

The OmpF measure is fitted, but only barely. Its training set is 45 molecules carrying a continuous relative permeability coefficient rather than a pass/fail label. With so few molecules a classifier faces two problems. First, the continuous values would have to be cut into two classes by hand, which means inventing a cut-off. Second, a five-fold cross-validation would leave only nine molecules in each test fold, so the resulting figure would not be reliable. This measure is therefore a written formula, in which the three features are standardised and added, with molecular size taking a minus sign while net charge and the entry primary amine take a plus sign. Only nine quantities are estimated from those 45 molecules, namely three means, three standard deviations and three signs, and these are then frozen rather than recomputed batch by batch.

The OmpF measure uses measured relative permeability coefficients scaled as OmpF/glycine = 100\%, from four published sources \citep{boi2025,ferreira2019,zahn2016,acosta2018}; the multi-porin measure uses direct liposome measurements, with the per-entry provenance in Note S4. Both porin measures are limited in training data and no further data could be found; this is a limitation of the work.

This measure uses three features: molecular weight, net charge at pH 7.4 and number of primary amines on sp3 carbon. It is a written formula rather than a fitted model, so no algorithm is used. The training set holds 45 molecules, and the label is a continuous permeability coefficient. The external set is the marketed 544.

\subsection{Multi-porin measure}

The other applies to OmpF, OmpC and PhoE together, and its training set aggregates the hydrophilic molecules measured on all three. Other porins exist in \textit{E. coli} but none has usable data; OmpC and PhoE have no more than ten molecules tested each, whereas OmpF has 45.

The multi-porin measure, for example, was built with both Logistic Regression and Random Forest. Logistic Regression was dropped, although its internal leave-one-out AUC was in fact the higher of the two, because it barely ordered the molecules between the 0.5 and 0.97 percentiles.

This measure uses Random Forest \citep{breiman2001}. It takes eight features: molecular weight, net charge at pH 7.4, number of primary amines on sp3 carbon, fraction of rotatable bonds, transverse dipole variation across conformers, minimum total dipole moment across conformers, zwitterionic group flag and zwitterionic charge flag. The training set holds 62 molecules, 34 of them positive and 28 negative, and each molecule is labelled as permeating or not. The external set is the marketed 544.

The multi-porin measure is an exception. Two of its features belong to the efflux measure. The reason remains to be investigated. One conjecture is that they carry some mechanism in porin permeation that is not yet understood. They are therefore kept.

\subsection{Lipid route measure}

The lipid route measure is built on the CO-ADD lpxC case-control screen. It does not use liposome data. This measure began with the same nine features as the others, and it was then reduced to five on the same grounds. Two features carry most of its weight, logP at pH 7.4 and fraction of rotatable bonds. Both are assigned to the outer-membrane barrier in their source papers \citep{cooper2018}. A high score therefore means that the outer membrane does not hold the molecule back. It does not mean that the molecule crosses by simple diffusion. The case-control compares two strains, and these two strains differ in their lipopolysaccharide layer. This measure therefore isolates that barrier, and not the porin pathway. The two routes are parallel, not nested.

This measure uses Random Forest. It takes five features: logP at pH 7.4, number of primary amines on sp3 carbon, fraction of rotatable bonds, transverse dipole variation across conformers and minimum total dipole moment across conformers. The training set holds 551 molecules, 353 of them positive and 198 negative, and each molecule is labelled as limited by the outer membrane or not. This measure has no external set for its own endpoint.

\subsection{Efflux measure}

The efflux measure is built on the CO-ADD tolC screen. It is then validated on El Zahed 1,462, of which 1,060 are pumped and 402 are not. Within this measure, logP at pH 7.4 accounts for more than 90\% of the importance. Even so, this work does not claim that the efflux measure is the same as lipophilicity alone. One corroboration comes from other data, in which efflux activity was observed only in compounds with a logP at or above -1.986 \citep{teelucksingh2022}. That is the same direction as here. However, that study covers only 52 compounds, so this point could be examined further if more data become available.

This measure uses Logistic Regression. It takes two features: logP at pH 7.4 and zwitterionic charge flag. The training set holds 933 molecules, 744 of them pumped and 189 not pumped. The external set is El Zahed 1,462.

\subsection{Activity measure}

The activity measure predicts whether a molecule inhibits the growth of \textit{E. coli} as a whole. That outcome is the joint result of all the steps above, so this measure works as a supplement to the other two tracks. Modeling activity from public data has been done before \citep{pogodin2019,dieguezsantana2022,zhou2025}. The measure itself is built on ChEMBL. A molecule counts as active when its MIC is 8 $\mu$g/mL or lower, following a published benchmark \citep{zhou2025}.

This measure uses XGBoost \citep{chentg2016}. It takes all nine features. The training set holds 16,237 molecules, 4,582 of them active and 11,655 inactive, with activity defined as an MIC of 8 $\mu$g/mL or lower. The external set is the marketed 544.

\subsection{Tool output}

\begin{table}[!t]
\caption{Structure of the 20-column output.\label{tab1}}
\centering
\begin{tabular}{lL{0.70\textwidth}}
\toprule
Position & Column \\
\midrule
1--3 & set / InChIKey / SMILES \\
4--6 & OmpF measure: score / percentile / decision cut \\
7--9 & multi-porin measure: score / percentile / decision cut \\
10--12 & lipid route measure: score / percentile / decision cut \\
13--14 & two self-promoted-uptake flags (aminoglycoside-like, lipopeptide-like) \\
15--16 & efflux measure: score / percentile (no decision cut) \\
17--19 & activity measure: score / percentile / enrichment call \\
20 & missing descriptors \\
\botrule
\end{tabular}
\end{table}

Table 1 gives, for each measure, the score of a molecule, its percentile against the CO-ADD reference set, and the decision cut where one exists. Two further columns state the possibility of the two kinds of self-promoted uptake, aminoglycoside-like and lipopeptide-like, marked as low, middle or high; both concern the permeation side only. Across the whole 173,925-row output, 102 molecules are marked middle or high, that is 0.06\%, so these two columns are an exception flag rather than a dimension in daily use. The remaining four columns are the source dataset, the SMILES string, its InChIKey, and the missing descriptors.

Only the mark and percentile of each measure are given; there is no total score. One example is given in Note S3. Ciprofloxacin receives a position on each of the five measures, and the five are never combined into one. It is shown as an illustration rather than as a random sample. The final cut-off is chosen by the user. The decision cut on the table is just a recommendation. It was set on the shape of its own curve rather than by a common rule, and no cut is claimed to be optimal. The efflux measure has no decision cut. Its curve is monotonic, so the percentile alone is readable, and the values quoted for it are reference points rather than cuts. The fewer molecules selected, the higher the precision, since precision falls as recall rises. The precision and recall of each measure across cut-offs are given in Note S2.

MechGate is implemented in Python and runs from the command line: the input is a single column of SMILES, and the output is the 20-column table described above. It depends only on open-source packages --- RDKit for feature calculation and conformer generation \citep{landrum2023}, scikit-learn for the random forest and logistic regression measures \citep{pedregosa2011}, and XGBoost for the activity measure \citep{chentg2016} --- and requires no commercial software.

\subsection{Statistical analysis}

Intervals for proportions are Wilson intervals \citep{wilson1927}, while intervals for paired differences come from 2,000 bootstrap resamples \citep{efron1979}. Correlations are reported as Spearman's rho \citep{spearman1904} and agreement as Cohen's kappa \citep{cohen1960}. Every readout is accompanied by a shuffled-label baseline.

Throughout, molecules that appear in the training set of any measure involved in a readout are removed, so the subset differs between readouts: 536, 533, 528, 529 and 469 for the five measures taken singly, and 515 where four measures are combined.

\subsection{Data collection}

All training, testing and validation data come from six sources.
1, OmpF data on 45 molecules, drawn from four published sources \citep{boi2025,ferreira2019,zahn2016,acosta2018}.
2, Porin data on 62 molecules, drawn from four articles measuring permeation through porins \citep{yoshimura1985,nikaido1983,sugawara2016,boi2025}.
3, CO-ADD (CO-ADD).
4, ChEMBL (ChEMBL).
5, El Zahed's data \citep{elzahed2021}.
6, the 544 marketed drug molecules from \citet{krishnan2025}.

 The selection steps applied to each source are given in Note S5.

\section{Results}

The five measures were first evaluated on the external marketed set. At their decision cuts, precision ranges from 0.789 to 0.873 and recall from 0.217 to 0.600 (Table 2). Each measure is evaluated on its own overlap-removed subset, and the five measures are trained on different endpoints; the rows are therefore not directly comparable. Figure 3 shows the same five measures across percentile cut-offs.

Within the training sets on which the measures were built, no decision cut applies, so performance is summarised by AUC instead. The multi-porin measure gives a leave-one-out AUC of 0.7731, its training set being too small for any other split, while the lipid route, efflux and activity measures give cross-validated AUCs of 0.7729, 0.7077 and 0.699. The OmpF measure is not trained at all and has no such readout.

\begin{figure}[!t]
\centering
\includegraphics[width=\textwidth]{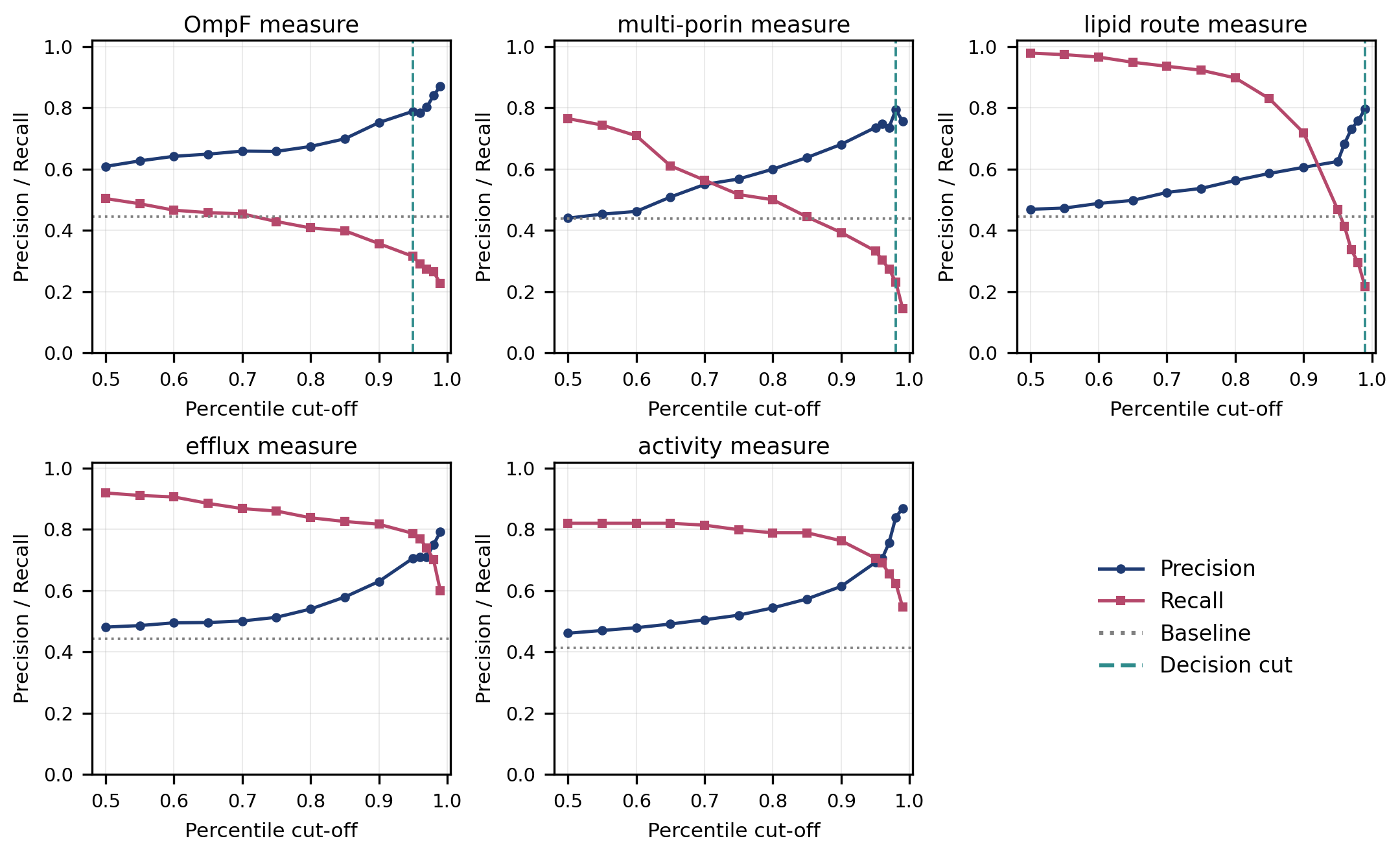}
\caption{Precision and recall of each measure across percentile cut-offs, on its own overlap-removed subset of the marketed reference set. The dotted line is the baseline of that subset; the dashed line marks the decision cut. The efflux and activity measures have no percentile cut and are shown without one.}\label{fig3}
\figalttext[Precision and recall against percentile cut-off]{Five panels, one per measure, each plotting precision and recall against the percentile cut-off. A dotted line marks the baseline of the subset and, where one exists, a dashed line marks the decision cut.}
{\footnotesize Alt text: Five panels, one per measure, each plotting precision and recall against the percentile cut-off. A dotted line marks the baseline of the subset and, where one exists, a dashed line marks the decision cut.}
\end{figure}

Alt text: Five panels, one per measure, each plotting precision and recall against the percentile cut-off. A dotted line marks the baseline of the subset and, where one exists, a dashed line marks the decision cut.

\begin{table}[!t]
\caption{Precision and recall of each measure at its decision cut, on its own overlap-removed subset of the marketed 544. The subsets differ between measures, so the rows are not comparable with one another. * The efflux measure has no decision cut; this value is a reference point.\label{tab2}}
\centering\resizebox{\textwidth}{!}{%
\begin{tabular}{llrrrrrr}
\toprule
Measure & Decision cut & n & Positives & Baseline & Calls & Precision & Recall \\
\midrule
OmpF & percentile $\ge$ 0.95 & 536 & 238 & 0.444 & 95 & 0.789 & 0.315 \\
multi-porin & percentile $\ge$ 0.98 & 533 & 234 & 0.439 & 68 & 0.794 & 0.231 \\
lipid route & percentile $\ge$ 0.99 & 528 & 235 & 0.445 & 64 & 0.797 & 0.217 \\
efflux & percentile $\ge$ 0.99 * & 529 & 235 & 0.444 & 178 & 0.792 & 0.600 \\
activity & score $\ge$ 0.5 & 469 & 194 & 0.414 & 126 & 0.873 & 0.567 \\
\botrule
\end{tabular}
}
\end{table}

The marks can nevertheless be combined, which makes them more interpretable. A molecule that is both permeable and efflux-avoidant is more likely to be effective in mechanistic terms. A molecule scoring well on either porin measure or on the lipid route should be treated as permeable. Precision improves in mechanism combination, literally (OmpF $\cup$ multi-porin $\cup$ lipid) $\cap$ efflux. In addition, the mechanism combination recovers molecules that the activity measure calls negative but which are in fact active, and the activity measure does the same in reverse; molecules called positive by both are highly precise. This holds when the activity measure is taken at the 0.95 percentile or tighter.

Intersections are reported on the external set only. For mechanism combination, literally (OmpF $\cup$ multi-porin $\cup$ lipid) $\cap$ efflux, the precision is 0.92 when each permeation measure sits at its own decision cut, at the 0.95, 0.98 and 0.99 percentiles respectively, and the efflux measure at the 0.99 percentile. Table 3 gives this intersection across seven efflux percentiles, with the three permeation measures held at their own decision cuts. Each permeation measure crossed with the efflux measure over both axes is given in Note S6. This table uses the subset of 515 molecules left after removing those in the training set of any of the four measures involved.

\begin{table}[!t]
\caption{Precision and recall of (OmpF $\cup$ multi-porin $\cup$ lipid) $\cap$ efflux across efflux percentiles, with the three permeation measures held at their decision cuts, on the 515 molecules left after removing those in the training set of any measure involved (positives 224, baseline 0.435).\label{tab3}}
\centering\resizebox{\textwidth}{!}{%
\begin{tabular}{lrrrrrrr}
\toprule
Efflux percentile & 0.50 & 0.70 & 0.80 & 0.90 & 0.95 & 0.98 & 0.99 \\
\midrule
n & 168 & 159 & 154 & 149 & 141 & 125 & 106 \\
Precision & 0.768 & 0.761 & 0.760 & 0.785 & 0.823 & 0.888 & 0.924 \\
Recall & 0.576 & 0.540 & 0.522 & 0.522 & 0.518 & 0.495 & 0.438 \\
\botrule
\end{tabular}
}
\end{table}

The 442 molecules are the original 544 with the training molecules of all five measures removed. Of these, 52 are called by both the mechanism combination and the activity measure and 51 are positive (0.981); 25 by the mechanism combination alone, 18 positive (0.720); and 62 by the activity measure alone, 47 positive (0.758), against a baseline of 0.396 on this subset (Note S11). Here the three permeation measures are taken at their own decision cut-offs and combined by union, the efflux measure at the 0.99 percentile, and the activity measure at its native threshold.

Three of the five measures do better than ranking by their single most important feature, while for the other two the point estimates are positive but the intervals include zero. The per-measure differences and their intervals are given in Note S7.

In a batch of over 3,000 molecules from \citet{kollen2025}, tightening the cut-off from the 50th to the 99th percentile reduces the fraction selected by the OmpF measure from 77.5\% to 11.1\%, by the multi-porin measure from 54.5\% to 3.6\%, and by the lipid measure from 80.6\% to 4.8\% (Note S8). This is shown in Figure 4.

\begin{figure}[!t]
\centering
\includegraphics[width=\textwidth]{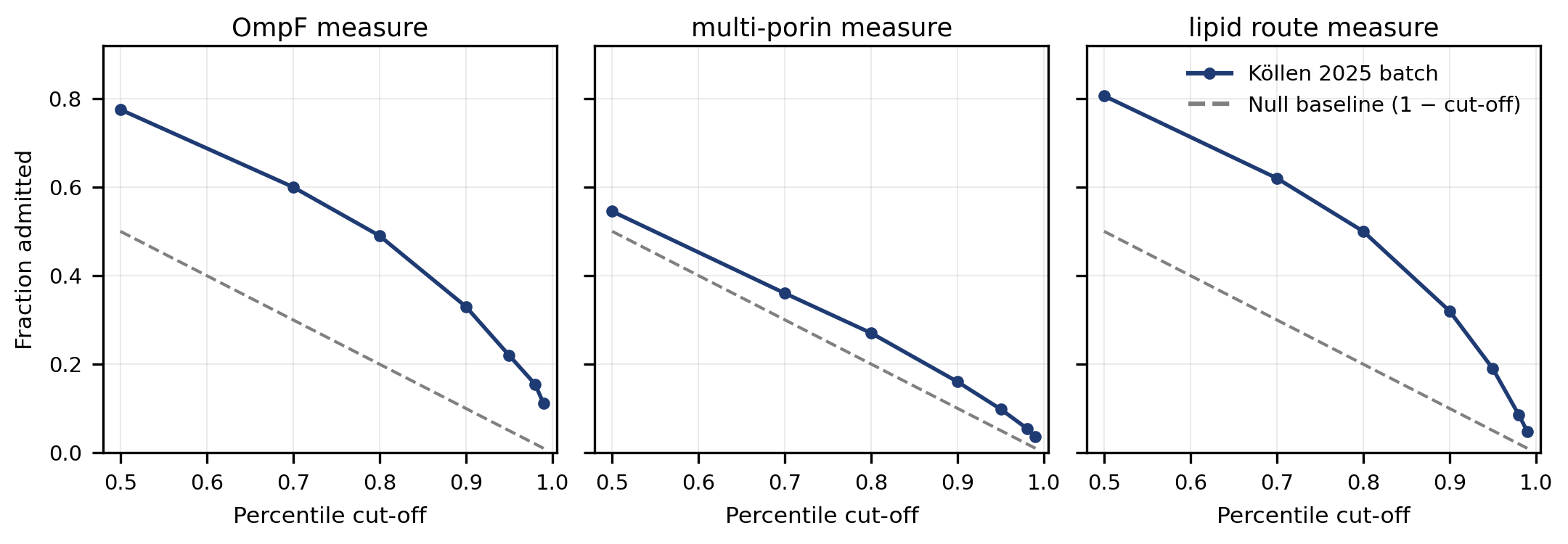}
\caption{Fraction of a generated batch still admitted as the cut-off tightens, on the three permeation measures. The dashed line is what a batch indistinguishable from the reference set would give.}\label{fig4}
\figalttext[Fraction of a generated batch still admitted]{Three panels, one per permeation measure, each plotting the fraction of a generated batch still admitted as the percentile cut-off tightens. A dashed line shows what a batch indistinguishable from the reference set would give.}
{\footnotesize Alt text: Three panels, one per permeation measure, each plotting the fraction of a generated batch still admitted as the percentile cut-off tightens. A dashed line shows what a batch indistinguishable from the reference set would give.}
\end{figure}

Alt text: Three panels, one per permeation measure, each plotting the fraction of a generated batch still admitted as the percentile cut-off tightens. A dashed line shows what a batch indistinguishable from the reference set would give.

The four source libraries are separable from one another pairwise using physicochemical features alone, with AUCs from 0.809 to 0.941, and none of the 18 confidence intervals crosses 0.5. Across ten source libraries, the within-library and cross-library rankings invert as a whole, and sample size does not account for the inversion. The pairwise AUCs are given in Note S9 and the cross-library rankings in Note S10.

\section{Discussion and conclusion}

\subsection{Discussion}

We first attempted to adopt features from existing works. Adding features without mechanistic grounds mainly makes the models score well within their own datasets, but that score does not transfer to other sets. Especially in the efflux measure, logP at pH 7.4 alone accounts for more than 90\% of the importance. An efflux measure with two mechanistically supported features therefore performs better than models carrying more features without such support.

One further point concerns the choice of measures. We failed to unify the permeation and efflux measures into one measure with a single total score, since no dataset carries both labels. Data limitation also caused the separation in permeation track. The available data on permeation cover three different endpoints: the OmpF permeability coefficient, direct liposome measurement, and the CO-ADD lpxC case-control. The first two measure passage through a porin; the third measures how strongly the outer membrane holds a molecule back. With different endpoints, these cannot be unified into one measure. The final tool therefore has five measures: two porin measures, one lipid route measure, one efflux measure and one activity measure.

\subsection{Limitations}

\subsubsection{External sets and validation}

No permeation-side measure has an external set that is both endpoint-matched and carries a negative class. Only the efflux measure has one, in El Zahed 1,462. El Zahed 650 is endpoint-matched but wholly positive, so only recall can be reported there (Note S12); the marketed 544 carries negatives but its label is activity. The liposome set (n = 59) and the eNTRy set (n = 361) do carry negatives, but the false-positive rate has not been computed on either.

The OmpF measure has no independent external unit and no independent negative evidence. After overlap removal n = 3. Its negative calls track molecular weight, but that sign is fixed negative by construction.

The lipid route measure has no external set for its own endpoint. Only a cross-validated out-of-fold estimate within its training set, so feature changes cannot be judged externally.

\subsubsection{What a negative call means}

A negative call means ``not detected'', not ``did not enter''. CO-ADD screens at a single concentration (CO-ADD); elsewhere negatives rest on signal below the limit of detection. Every negative class here carries this, training sets included.

Two readouts are unexplained. The activity measure agrees with efflux (kappa 0.514) far more than with permeation. And on shuffled labels 28 of 56 cells exclude zero, so the gap measure carries an offset; the five cells quoted all include zero.

\subsubsection{Coverage and definitions}

The two porin-side measures use different definitions, and neither describes the porin protein. One takes OmpF alone, the other the best of OmpF, OmpC and PhoE. Among 71 comparable entries 3 reverse rank, 18 of them from one batch \citep{mahendran2010}.

The mechanism side does not cover every route in. Of the 62 molecules called by the activity measure alone, 47 are positive.

\subsubsection{The basis of individual measures}

The 3D features are expensive and not open. The original size feature needed 2.56 days over 74k molecules and was replaced by molecular weight; verifying it independently would require commercial software, which was not used here.

Retention in Le Goff is not intracellular concentration. It is the sum of uptake, membrane insertion and surface adsorption \citep{legoff2025}, and the original does not say which barrier it mainly reflects.

The activity threshold was read off the distribution. MIC $\le$ 8 $\mu$g/mL, following a published benchmark \citep{zhou2025}, with no independent grounding.

The efflux measure does not resolve pump type. Thirty-odd pumps, one aggregate readout. Single-pump data exist \citep{teelucksingh2022} but have a different endpoint. Molecules entering by self-promoted uptake \citep{hancock1991} bypass both porin and lipid routes and are flagged, not scored.

\subsection{Conclusion}

In this study, MechGate is built to screen small drug molecules against \textit{E. coli} by mechanism. It splits entry into three tracks and five measures: two porin measures, one lipid route measure, one efflux measure and one activity measure. Each measure is built on a published \textit{E. coli} endpoint, and each feature is taken from a paper that states its mechanism. On the external marketed set the five measures reach a precision of 0.789 to 0.873 and a recall of 0.217 to 0.600 at their decision cuts. Taken in mechanism combination the precision rises to 0.92. The five readings are reported side by side, and no total score is formed.

The construction itself is not specific to \textit{E. coli}. What ties it to this organism is the data: the \textit{E. coli} endpoints it is built on carry no structural parameter of any porin or pump. For another species the same procedure could be followed, but each measure would have to be rebuilt on that species' own measured data.

\section{Author biographies}

Lang Qin is a graduate student in the MComp Applied AI programme at the College of Computing and Data Science, Nanyang Technological University. His work concerns computational antibiotic discovery for Gram-negative bacteria.

Jagath C Rajapakse is a professor at the College of Computing and Data Science, Nanyang Technological University. His research covers machine learning, computational biology and biomedical data analysis.

\section{Acknowledgements}

We thank Dr Wang Zechen for his comments on the manuscript.

\section{Data availability}

Source code, the frozen parameters of the five measures, and the test data needed to reproduce every readout, table and figure reported here are freely available at https://github.com/qinlang0008-blip/mechgate under the GPL-2.0 licence, with no registration required; the submitted version and its test data are archived at https://doi.org/10.5281/zenodo.23431905. Installation and a worked example are described in the repository README.
\nocite{coadd,chembl,bindingdb}
\bibliographystyle{oup-abbrvnat}
\bibliography{reference}



\section*{Supplementary material}

# Supplementary Information

Notes are numbered in the order they are first cited in the main text; tables within them are numbered sequentially.

Conventions. All readouts on the marketed reference set are taken after removing the molecules that appear in the training set of any measure involved, so the subset differs between tables; each table states its own n, positives, negatives and baseline. Tables with different denominators are not comparable with one another. All proportions are given to three decimal places.

## Note S1 · Feature definitions and the seven removed features

Main-text location: Features and algorithms. Not on disk; contents taken from the SI Inventory, Section 5. This is a prose section, not a table.

Feature definitions. Each feature is listed below with the column name it carries in the tool output, its definition, and the point at which the implementation departs from a published default.

• molecular weight (mol_size) — molecular weight in Da, from RDKit Descriptors.MolWt, used as a linear value. Krishnamoorthy 2017 plots sizes as log(MW); on this set the log and the linear value rank the molecules identically, so the linear value is used.

• net charge at pH 7.4 (net_charge_v2) — the integer net formal charge of the pH 7.4 microspecies generated by Open Babel, then corrected by two SMARTS rules: minus one for each acylated or heteroaryl sulfonamide, plus one for each formamidine. With no rule applied it equals the uncorrected Open Babel value.

• number of primary amines on sp3 carbon (n_prim_amine_entry) — the count of primary amines bonded to an sp3 carbon, matched by [$([N;H2;X3][CX4]),$([N;H3;X4+][CX4])] in both the neutral and the protonated form. This is the eNTRy definition, so aromatic amines, amino heterocycles and guanidines are excluded. It is not the broad version listed above.

• logP at pH 7.4 (slogp) — the Wildman and Crippen logP, from RDKit Crippen.MolLogP, computed on the pH 7.4 microspecies generated by Open Babel. There is no normalisation and no sign change.

• fraction of rotatable bonds (b_rotR) — the number of non-ring single bonds whose two atoms each have more than one heavy-atom neighbour, divided by the total number of heavy-atom bonds. This is the MOE and Cooper 2018 convention, which does not exclude amide bonds, and it is therefore not the RDKit rotatable-bond count.

• transverse dipole variation across conformers (dxy_cv) — for each conformer the dipole is projected onto the plane perpendicular to the principal axis of least inertia, which gives a transverse dipole; the feature is the coefficient of variation of that quantity over the conformer ensemble, that is the sample standard deviation divided by the mean. The ensemble holds 20 conformers from ETKDGv3, optimised with MMFF94s, with Gasteiger charges and the dipole origin at the centre of mass. The two directions are relative to the molecular axis, not Cartesian axes. The quantity follows Acosta-Gutiérrez 2018 and the choice of a variation statistic over a mean follows Boi 2025.

• minimum total dipole moment across conformers (dtotal_min) — the smallest total dipole moment, in Debye, over the same conformer ensemble. It is the minimum across the ensemble, and not the value at the lowest-energy conformer, which is held in a separate column.

• zwitterionic group flag (zwit_group) — one when the molecule carries at least one dissociable acidic group and at least one protonatable basic group, matched on the input structure. It does not read the net charge or the protonation state.

• zwitterionic charge flag (zwit_charge) — one when the net charge at pH 7.4 is exactly zero and a positive site and a negative site are both present. The two flags can disagree, which is expected, and both are kept.

Seven removed features. The grounds for each removal are given below.

• fsp3 / hwi — removing them had no detectable adverse effect

• glob_min — the accessible sources give only a definition and a threshold

• rot_rigid_rdkit — Richter, R. 2020 uses the relative abundance of unsaturated bonds, in a potato-starch hydrogel rather than a bacterial outer membrane; it is therefore not the same quantity as the one used here

• logp_ow — the mechanistic accounts disagree (the evidence in Cooper lies on the efflux side, while Boi 2025 reports that logP is unrelated to porin permeation)

• n_prim_amine, broad version — the implementation does not match the original definition

## Note S2 · Precision and recall of each measure across percentile cut-offs

Each measure is shown on its own overlap-removed subset. The subsets differ, so the five panels are not comparable with one another. The activity measure has an extra column for its native threshold, which is a score rather than a percentile.

Table S1 · OmpF measure — n = 536 (238 positive / 298 negative), baseline 0.444

| Cut-off | 0.50 | 0.55 | 0.60 | 0.65 | 0.70 | 0.75 | 0.80 |
|---|---|---|---|---|---|---|---|
| n | 197 | 185 | 173 | 168 | 164 | 155 | 144 |
| Precision | 0.609 | 0.627 | 0.642 | 0.649 | 0.659 | 0.658 | 0.674 |
| Recall | 0.504 | 0.487 | 0.466 | 0.458 | 0.454 | 0.429 | 0.408 |

| Cut-off | 0.85 | 0.90 | 0.95 | 0.96 | 0.97 | 0.98 | 0.99 |
|---|---|---|---|---|---|---|---|
| n | 136 | 113 | 95 | 88 | 81 | 75 | 62 |
| Precision | 0.699 | 0.752 | 0.789 | 0.784 | 0.802 | 0.840 | 0.871 |
| Recall | 0.399 | 0.357 | 0.315 | 0.290 | 0.273 | 0.265 | 0.227 |

Table S2 · multi-porin measure — n = 533 (234 positive / 299 negative), baseline 0.439

| Cut-off | 0.50 | 0.55 | 0.60 | 0.65 | 0.70 | 0.75 | 0.80 |
|---|---|---|---|---|---|---|---|

| | | | | | | | |
|---|---|---|---|---|---|---|---|
| n | 407 | 384 | 359 | 281 | 240 | 213 | 195 |
| Precision | 0.440 | 0.453 | 0.462 | 0.509 | 0.550 | 0.568 | 0.600 |
| Recall | 0.765 | 0.744 | 0.709 | 0.611 | 0.564 | 0.517 | 0.500 |

| **Cut-off** | **0.85** | **0.90** | **0.95** | **0.96** | **0.97** | **0.98** | **0.99** |
|---|---|---|---|---|---|---|---|
| n | 163 | 135 | 106 | 95 | 87 | 68 | 45 |
| Precision | 0.638 | 0.681 | 0.736 | 0.747 | 0.736 | 0.794 | 0.756 |
| Recall | 0.444 | 0.393 | 0.333 | 0.303 | 0.274 | 0.231 | 0.145 |

Table S3 · lipid route measure — n = 528 (235 positive / 293 negative), baseline 0.445

| **Cut-off** | **0.50** | **0.55** | **0.60** | **0.65** | **0.70** | **0.75** | **0.80** |
|---|---|---|---|---|---|---|---|
| n | 490 | 484 | 465 | 448 | 420 | 404 | 375 |
| Precision | 0.469 | 0.473 | 0.488 | 0.498 | 0.524 | 0.537 | 0.563 |
| Recall | 0.979 | 0.974 | 0.966 | 0.949 | 0.936 | 0.923 | 0.898 |

| **Cut-off** | **0.85** | **0.90** | **0.95** | **0.96** | **0.97** | **0.98** | **0.99** |
|---|---|---|---|---|---|---|---|
| n | 333 | 279 | 176 | 142 | 108 | 91 | 64 |
| Precision | 0.586 | 0.606 | 0.625 | 0.683 | 0.731 | 0.758 | 0.797 |
| Recall | 0.830 | 0.719 | 0.468 | 0.413 | 0.336 | 0.294 | 0.217 |

Table S4 · efflux measure — n = 529 (235 positive / 294 negative), baseline 0.444

| **Cut-off** | **0.50** | **0.55** | **0.60** | **0.65** | **0.70** | **0.75** | **0.80** |
|---|---|---|---|---|---|---|---|
| n | 449 | 440 | 430 | 419 | 407 | 394 | 365 |
| Precision | 0.481 | 0.486 | 0.495 | 0.496 | 0.501 | 0.513 | 0.540 |
| Recall | 0.919 | 0.911 | 0.906 | 0.885 | 0.868 | 0.860 | 0.838 |

| **Cut-off** | **0.85** | **0.90** | **0.95** | **0.96** | **0.97** | **0.98** | **0.99** |
|---|---|---|---|---|---|---|---|
| n | 335 | 305 | 262 | 255 | 245 | 220 | 178 |
| Precision | 0.579 | 0.630 | 0.706 | 0.710 | 0.710 | 0.750 | 0.792 |
| Recall | 0.826 | 0.817 | 0.787 | 0.770 | 0.740 | 0.702 | 0.600 |

Table S5 · activity measure — n = 469 (194 positive / 275 negative), baseline 0.414

| **Cut-off** | **0.50** | **0.55** | **0.60** | **0.65** | **0.70** | **0.75** | **0.80** |
|---|---|---|---|---|---|---|---|
| n | 345 | 338 | 332 | 324 | 313 | 298 | 281 |
| Precision | 0.461 | 0.470 | 0.479 | 0.491 | 0.505 | 0.520 | 0.544 |
| Recall | 0.820 | 0.820 | 0.820 | 0.820 | 0.814 | 0.799 | 0.789 |

| **Cut-off** | **0.85** | **0.90** | **0.95** | **0.96** | **0.97** | **0.98** | **score ≥ 0.5** | **0.99** |
|---|---|---|---|---|---|---|---|---|
| n | 267 | 241 | 198 | 190 | 168 | 144 | 126 | 122 |
| Precision | 0.573 | 0.614 | 0.692 | 0.705 | 0.756 | 0.840 | 0.873 | 0.869 |
| Recall | 0.789 | 0.763 | 0.706 | 0.691 | 0.655 | 0.624 | 0.567 | 0.546 |

## Note S3 · The 20-column output for one molecule

Main-text location: Tool output. The example is Ciprofloxacin; each measure gives a score and a percentile against the CO-ADD reference set, and the five are never combined into one.

Table S6 · The 20-column output for one molecule, Ciprofloxacin

| set | InChIKey | SMILES | OmpF score | OmpF percentile |
|---|---|---|---|---|
| marketed 544 | MYSWGUAQZAJSOK-UHFFFAOYSA-N | O=C(O)c1cn(C2CC2)c2cc(N3CCNCC3)c(F)cc2c1=O | -0.494 | 0.677 |
| **OmpF cut** | **multi-porin score** | **multi-porin percentile** | **multi-porin cut** | **lipid score** |
| low | 0.807 | 0.996 | high | 0.160 |
| **lipid percentile** | **lipid cut** | **aminoglycoside-like** | **lipopeptide-like** | **efflux score** |
| 0.990 | high | low | low | 0.207 |
| **efflux percentile** | **activity score** | **activity percentile** | **activity call** | **missing descriptors** |
| 0.994 | 0.779 | 1.000 | selected | (none) |

## Note S4 · Detailed composition and per-entry provenance of each training set

Main-text location: Data collection. Not on disk; contents taken from SI Inventory, Section 5.

• Composition of the 48 entries of the OmpF measure, from four sources: 35 from Boi 2025 + 8 from Ferreira 2019 + 3 from Zahn 2016 + 2 from Acosta-Gutiérrez 2018. Source file section34/scale_A_v2.tsv (md5 f8515728b9389fe3a36f8a81a157a79b, skiprows=6); the 48 InChIKeys are distinct. Entries are counted by the source they were recorded from, which is not necessarily the laboratory that measured them.

• The 48 entries and the 45 used for standardisation: three sugars (glucose, sucrose, maltose) are removed by the scope rule, and the means and standard deviations of the three features are frozen on the remaining 45. By source, those 45 are 35 from Boi 2025, 7 from Ferreira 2019, 1 from Zahn 2016 and 2 from Acosta-Gutiérrez 2018. The measure fits no weights.

• Must be stated: all four sources normalise to OmpF with glycine = 100%.

• Difference between the efflux training set (933) and the full case-control set (965): an earlier design required four features to be simultaneously non-empty, whereas the current measure uses only two; the 32-molecule difference was not retrained.

• Composition of the 3,809 entries of the third-level stratification: 250 must-permeate molecules / 253 records + 3,531 measured non-retained + 25 removed by the pre-set criterion but kept traceable.

• Per-entry breakdown of the 27 ORD units: 9 units with $n < 4$ for which a correlation cannot be computed; only T04×OmpC and T04×OmpF have zero overlap, with just 3 molecules each.

## Note S5 · Selection steps applied to each source

Main-text location: Data collection. Each source was filtered before use; the steps and the resulting counts are given here.

ChEMBL, for the activity measure. Four steps were applied to 92,941 raw records.

Table S7 · ChEMBL, four filtering steps

| Step | Rule | Remaining |
|---|---|---|
| 1 | Strain restricted to ATCC 25922, the wild type used by CO-ADD | 21,853 records |
| 2 | Units convertible to µg/mL (µg/mL directly, nM converted by molecular weight) | 21,132 records |
| 3 | Records merged to one value per molecule, taking the median | 17,680 molecules |
| 4 | Called active at MIC ≤ 8 µg/mL; `>` counted as inactive, `<` and `≤` as active | 5,133 active / 12,547 inactive |

After features were computed and molecules whose 3D conformer failed were dropped, the training set is 16,237 molecules (4,582 positive / 11,655 negative). The threshold of 8 µg/mL was read off the distribution and has no independent grounding.

CO-ADD, for the lipid route and efflux measures. Of 74,176 molecules, 5 were removed by the inclusion criteria and a further 1,017 because features could not be computed, leaving 73,154. This set is the reference against which all percentiles are taken; it is not a training set.

Two case-control sets were built from the same screen, differing only in the mutant strain. In each, molecules active on the mutant were selected first, and the wild-type reading then decided the label:

Table S8 · The two CO-ADD case-control sets

| Quantity | Definition |
|---|---|
| Sample set | molecules with mutant inhibition above mean + 4.0 SD |
| Label | `limited` if wild-type inhibition is at or below mean + 4.0 SD |
| lipid route | *lpxC* mutant; 551 molecules after feature filtering (353 positive / 198 negative) |
| efflux | *tolC* mutant; 933 molecules after feature filtering (744 positive / 189 negative) |

Inclusion criteria. Two rules were applied, each removing a molecule only if every condition held.

Table S9 · The two inclusion rules

| Rule | Conditions |
|---|---|
| Sugars | at least one ring oxygen of sugar type; at least three alcohol hydroxyls; no ionisable group; no β-lactam; no aromatic ring; no non-sugar ring |
| Oligopeptides | no β-lactam; at least one backbone peptide bond; at least one free α-amino group; at least one free carboxyl; no ring; at most four peptide bonds |

A molecule whose structure could not be parsed was kept rather than removed.

Krishnan 2025, for the marketed reference set. From 559 known antibacterials, one duplicate was removed by InChIKey and 14 molecules were dropped because their 3D conformer could not be generated, leaving 544. The Gram-negative label is taken from the source table without modification (244 positive / 300 negative).

El Zahed, for the efflux measure. Two sets are used: 650 molecules carrying a true permeation label, all positive; and 1,462 molecules labelled pumped or not pumped (1,060 / 402), of which 1,343 have both efflux features and can be scored.

### Complete inventory of sources for the porin permeation data

Main-text location: Data collection. Not on disk; contents taken from SI Inventory, Section 5.

• Step-by-step counts for the four ChEMBL filtering steps: 92,941 → 21,853 → 21,132 → 17,680 → 5,133 active / 12,547 inactive → 4,850 after removing overlap → 4,582 with a complete feature set

• Why SMOTE synthesis and the clinical-antibiotic list do not fill the gap: of the 738 molecules in Gurvic, 183 are themselves the left-hand block of the third-level stratification

• Verbatim comparison showing that Accum_68 of AutoMolDesigner and table4 of AIbiotics both have 36/32 positives and negatives

• Case-control construction for the lipid route measure and the efflux measure (inhib_lpxc > mean + 4σ and inhib_wt ≤ mean + 4σ, etc.)

### Full conditions and implementation of the inclusion criteria

Main-text location: Data collection. Not on disk; contents taken from SI Inventory, Section 5. This is a prose section, not a table.

• The six conditions of Rule S: ≥1 sugar ring; ≥3 alcoholic hydroxyls; 0 strongly ionisable groups; no β-lactam ring; 0 aromatic rings; 0 non-sugar rings

• The veto clause of Rule P and its rationale: any molecule containing a β-lactam ring is never removed — otherwise the side chain of ampicillin and similar compounds would satisfy the peptide-bond criterion

• The three hard acceptance checks, item by item: fosfomycin retained / linezolid retained and not by virtue of the "neutral" condition alone / 0 of the 43 β-lactam entries removed

• Quantified effect of adding condition S6: among 3,553 candidates, removals fell from 24 to 1 (0.03%)

• Group A: 48 → 45 (glucose, sucrose and maltose removed)

• Implementation file: s34l_scope_filter.py

## Note S6 · Each permeation measure crossed with the efflux measure

Every cell gives precision / n / recall. The permeation percentile runs down the rows and the efflux percentile across the columns. Each panel is on its own overlap-removed subset, formed by removing the training sets of both measures involved; the three denominators differ (521, 518, 527) and the panels are not comparable with one another.

Table S10 · OmpF measure ∩ efflux measure — n = 521 (229 positive / 292 negative), baseline 0.440

| Permeation \ Efflux | E ≥ 0.50 | E ≥ 0.70 | E ≥ 0.80 | E ≥ 0.90 | E ≥ 0.95 | E ≥ 0.98 | E ≥ 0.99 |
|---|---|---|---|---|---|---|---|
| **P ≥ 0.50** | 0.584 / 173 / 0.441 | 0.578 / 161 / 0.406 | 0.605 / 147 / 0.389 | 0.659 / 135 / 0.389 | 0.727 / 121 / 0.384 | 0.804 / 107 / 0.376 | 0.900 / 90 / 0.354 |
| **P ≥ 0.70** | 0.637 / 146 / 0.406 | 0.630 / 135 / 0.371 | 0.633 / 128 / 0.354 | 0.669 / 121 / 0.354 | 0.734 / 109 / 0.349 | 0.796 / 98 / 0.341 | 0.901 / 81 / 0.319 |
| **P ≥ 0.80** | 0.656 / 128 / 0.367 | 0.650 / 117 / 0.332 | 0.649 / 111 / 0.314 | 0.686 / 105 / 0.314 | 0.740 / 96 / 0.310 | 0.793 / 87 / 0.301 | 0.890 / 73 / 0.284 |
| **P ≥ 0.90** | 0.735 / 102 / 0.328 | 0.723 / 94 / 0.297 | 0.719 / 89 / 0.279 | 0.744 / 86 / 0.279 | 0.780 / 82 / 0.279 | 0.818 / 77 / 0.275 | 0.909 / 66 / 0.262 |
| **P ≥ 0.95** | 0.784 / 88 / 0.301 | 0.765 / 81 / 0.271 | 0.763 / 76 / 0.253 | 0.795 / 73 / 0.253 | 0.829 / 70 / 0.253 | 0.864 / 66 / 0.249 | 0.917 / 60 / 0.240 |
| **P ≥ 0.98** | 0.838 / 74 / 0.271 | 0.826 / 69 / 0.249 | 0.818 / 66 / 0.236 | 0.844 / 64 / 0.236 | 0.885 / 61 / 0.236 | 0.915 / 59 / 0.236 | 0.945 / 55 / 0.227 |
| **P ≥ 0.99** | 0.871 / 62 / 0.236 | 0.871 / 62 / 0.236 | 0.867 / 60 / 0.227 | 0.881 / 59 / 0.227 | 0.912 / 57 / 0.227 | 0.945 / 55 / 0.227 | 0.943 / 53 / 0.218 |

Table S11 · multi-porin measure ∩ efflux measure — n = 518 (225 positive / 293 negative), baseline 0.434

| Permeation \ Efflux | E ≥ 0.50 | E ≥ 0.70 | E ≥ 0.80 | E ≥ 0.90 | E ≥ 0.95 | E ≥ 0.98 | E ≥ 0.99 |
|---|---|---|---|---|---|---|---|
| **P ≥ 0.50** | 0.445 / 346 / 0.684 | 0.457 / 317 / 0.644 | 0.478 / 295 / 0.627 | 0.554 / 249 / 0.613 | 0.641 / 209 / 0.596 | 0.701 / 177 / 0.551 | 0.762 / 151 / 0.511 |
| **P ≥ 0.70** | 0.531 / 207 / 0.489 | 0.534 / 193 / 0.458 | 0.546 / 183 / 0.444 | 0.594 / 165 / 0.436 | 0.660 / 144 / 0.422 | 0.725 / 120 / 0.387 | 0.827 / 98 / 0.360 |
| **P ≥ 0.80** | 0.568 / 169 / 0.427 | 0.570 / 158 / 0.400 | 0.584 / 149 / 0.387 | 0.642 / 134 / 0.382 | 0.712 / 118 / 0.373 | 0.792 / 96 / 0.338 | 0.910 / 78 / 0.316 |
| **P ≥ 0.90** | 0.658 / 111 / 0.324 | 0.673 / 107 / 0.320 | 0.686 / 105 / 0.320 | 0.720 / 100 / 0.320 | 0.769 / 91 / 0.311 | 0.829 / 76 / 0.280 | 0.935 / 62 / 0.258 |
| **P ≥ 0.95** | 0.701 / 87 / 0.271 | 0.714 / 84 / 0.267 | 0.723 / 83 / 0.267 | 0.759 / 79 / 0.267 | 0.773 / 75 / 0.258 | 0.841 / 63 / 0.236 | 0.960 / 50 / 0.213 |
| **P ≥ 0.98** | 0.772 / 57 / 0.196 | 0.782 / 55 / 0.191 | 0.782 / 55 / 0.191 | 0.811 / 53 / 0.191 | 0.827 / 52 / 0.191 | 0.913 / 46 / 0.187 | 0.950 / 40 / 0.169 |

| | | | | | | | |
|---|---|---|---|---|---|---|---|
| **P ≥ 0.99** | 0.725 / 40 / 0.129 | 0.737 / 38 / 0.124 | 0.737 / 38 / 0.124 | 0.757 / 37 / 0.124 | 0.778 / 36 / 0.124 | 0.871 / 31 / 0.120 | 0.923 / 26 / 0.107 |

Table S12 · lipid route measure ∩ efflux measure — n = 527 (235 positive / 292 negative), baseline 0.446

| Permeation \ Efflux | E ≥ 0.50 | E ≥ 0.70 | E ≥ 0.80 | E ≥ 0.90 | E ≥ 0.95 | E ≥ 0.98 | E ≥ 0.99 |
|---|---|---|---|---|---|---|---|
| **P ≥ 0.50** | 0.513 / 413 / 0.902 | 0.534 / 380 / 0.864 | 0.563 / 350 / 0.838 | 0.653 / 294 / 0.817 | 0.728 / 254 / 0.787 | 0.767 / 215 / 0.702 | 0.797 / 177 / 0.600 |
| **P ≥ 0.70** | 0.583 / 350 / 0.868 | 0.590 / 334 / 0.838 | 0.606 / 317 / 0.817 | 0.663 / 285 / 0.804 | 0.743 / 245 / 0.774 | 0.781 / 210 / 0.698 | 0.797 / 177 / 0.600 |
| **P ≥ 0.80** | 0.619 / 320 / 0.843 | 0.627 / 306 / 0.817 | 0.632 / 296 / 0.796 | 0.669 / 275 / 0.783 | 0.748 / 238 / 0.757 | 0.784 / 204 / 0.681 | 0.801 / 171 / 0.583 |
| **P ≥ 0.90** | 0.669 / 248 / 0.706 | 0.678 / 239 / 0.689 | 0.682 / 233 / 0.677 | 0.693 / 228 / 0.672 | 0.782 / 197 / 0.655 | 0.818 / 170 / 0.591 | 0.832 / 143 / 0.506 |
| **P ≥ 0.95** | 0.684 / 158 / 0.460 | 0.684 / 152 / 0.443 | 0.682 / 151 / 0.438 | 0.687 / 150 / 0.438 | 0.783 / 129 / 0.430 | 0.840 / 106 / 0.379 | 0.847 / 85 / 0.306 |
| **P ≥ 0.98** | 0.761 / 88 / 0.285 | 0.767 / 86 / 0.281 | 0.767 / 86 / 0.281 | 0.776 / 85 / 0.281 | 0.844 / 77 / 0.277 | 0.922 / 64 / 0.251 | 0.936 / 47 / 0.187 |
| **P ≥ 0.99** | 0.790 / 62 / 0.209 | 0.790 / 62 / 0.209 | 0.790 / 62 / 0.209 | 0.790 / 62 / 0.209 | 0.828 / 58 / 0.204 | 0.917 / 48 / 0.187 | 0.917 / 36 / 0.140 |

## Note S7 · Each measure against ranking by its single most important feature

Each measure is compared with ranking the same molecules by its single most important feature and taking the same number of positive calls. The direction of the feature is set by the sign of its Spearman correlation with that measure's percentile, not chosen by hand; all five came out ascending. Intervals are from 2,000 paired bootstrap resamples on the measure's own overlap-removed subset.

Table S13 · Each measure against its single most important feature

| Measure | Feature | n | Precision of the measure | Precision of the feature alone | Difference | 95% interval | Includes zero |
|---|---|---|---|---|---|---|---|
| OmpF measure | mol_size | 536 | 0.789 | 0.453 | +0.337 | +0.240 to +0.433 | no |
| multi-porin measure | mol_size | 533 | 0.794 | 0.338 | +0.456 | +0.320 to +0.593 | no |
| lipid route measure | slogp | 528 | 0.797 | 0.766 | +0.031 | −0.118 to +0.174 | yes |
| efflux measure | slogp | 529 | 0.792 | 0.770 | +0.023 | −0.017 to +0.063 | yes |

| | | | | | | | |
|---|---|---|---|---|---|---|---|
| activity measure | slogp | 469 | 0.873 | 0.746 | +0.127 | +0.051 to +0.206 | no |

## Note S8 · Admitted fraction by batch across seven cut-offs

Main-text location: Results. Source file: cut7_admit_v10.tsv(md5 fc0945033380f45b5126ade83a937ac5, skiprows=0; main block only, 105 rows = 5 batches × 3 measures × 7 cut-offs).

Table note: on the reference set, the admitted fraction at each cut-off equals 1 − cut-off, i.e. the shuffled-label baseline.

Cell = admitted fraction (admitted/n); the last row of each table gives the baseline for that cut-off.

Table S14 · the OmpF measure

| **Batch \ cut-off** | **0.5** | **0.7** | **0.8** | **0.9** | **0.95** | **0.98** | **0.99** |
|---|---|---|---|---|---|---|---|
| **① AIbiotics 3,395** | 0.7752 (2631/3394) | 0.6196 (2103/3394) | 0.5003 (1698/3394) | 0.3300 (1120/3394) | 0.2275 (772/3394) | 0.1538 (522/3394) | 0.1108 (376/3394) |
| **② AMD S13 9,344** | 0.5309 (4961/9344) | 0.3937 (3679/9344) | 0.3166 (2958/9344) | 0.2178 (2035/9344) | 0.1487 (1389/9344) | 0.0944 (882/9344) | 0.0629 (588/9344) |
| **③ Marketed set, 544** | 0.3097 (153/494) | 0.2368 (117/494) | 0.1984 (98/494) | 0.1377 (68/494) | 0.1012 (50/494) | 0.0607 (30/494) | 0.0364 (18/494) |
| **④ Collins, eight batches** | 0.1352 (21622/159979) | 0.0805 (12874/159979) | 0.0585 (9363/159979) | 0.0358 (5720/159979) | 0.0218 (3487/159979 ) | 0.0116 (1857/159979) | 0.0066 (1059/159979) |
| **⑤ El Zahed 663** | 0.7443 (492/661) | 0.5053 (334/661) | 0.3843 (254/661) | 0.2405 (159/661) | 0.1377 (91/661) | 0.0530 (35/661) | 0.0272 (18/661) |
| *Baseline (= 1 − cut-off)* | *0.50* | *0.30* | *0.20* | *0.10* | *0.05* | *0.02* | *0.01* |

Table S15 · the multi-porin measure

| **Batch \ cut-off** | **0.5** | **0.7** | **0.8** | **0.9** | **0.95** | **0.98** | **0.99** |
|---|---|---|---|---|---|---|---|
| **① AIbiotics 3,395** | 0.5445 (1848/3394) | 0.2655 (901/3394) | 0.1632 (554/3394) | 0.0978 (332/3394) | 0.0766 (260/3394) | 0.0530 (180/3394) | 0.0386 (131/3394) |
| **② AMD S13 9,344** | 0.5686 (5313/9344) | 0.3370 (3149/9344) | 0.2497 (2333/9344) | 0.1600 (1495/9344) | 0.1055 (986/9344) | 0.0582 (544/9344) | 0.0378 (353/9344) |
| **③ Marketed set, 544** | 0.7429 (367/494) | 0.4332 (214/494) | 0.3462 (171/494) | 0.2449 (121/494) | 0.1903 (94/494) | 0.1296 (64/494) | 0.0931 (46/494) |
| **④ Collins, eight batches** | 0.5608 (89006/158711) | 0.4127 (65497/158711) | 0.3267 (51850/158711 ) | 0.1901 (30165/158711) | 0.1015 (16104/158711) | 0.0463 (7355/158711) | 0.0245 (3884/158711) |
| **⑤ El Zahed 663** | 0.5633 (365/648) | 0.3673 (238/648) | 0.2485 (161/648) | 0.1312 (85/648) | 0.0833 (54/648) | 0.0324 (21/648) | 0.0154 (10/648) |

| | | | | | | | |
|---|---|---|---|---|---|---|---|
| *Baseline (= 1 − cut-off)* | *0.50* | *0.30* | *0.20* | *0.10* | *0.05* | *0.02* | *0.01* |

Table S16 · the lipid route measure

| Batch \ cut-off | 0.5 | 0.7 | 0.8 | 0.9 | 0.95 | 0.98 | 0.99 |
|---|---|---|---|---|---|---|---|
| **① AIbiotics 3,395** | 0.8058 (2735/3394) | 0.6111 (2074/3394) | 0.4859 (1649/3394) | 0.3111 (1056/3394) | 0.1836 (623/3394) | 0.0831 (282/3394) | 0.0486 (165/3394) |
| **② AMD S13 9,344** | 0.7535 (7041/9344) | 0.5244 (4900/9344) | 0.3724 (3480/9344) | 0.2021 (1888/9344) | 0.1138 (1063/9344) | 0.0552 (516/9344) | 0.0305 (285/9344) |
| **③ Marketed set, 544** | 0.9170 (453/494) | 0.7753 (383/494) | 0.6842 (338/494) | 0.5000 (247/494) | 0.3381 (167/494) | 0.1964 (97/494) | 0.1437 (71/494) |
| **④ Collins, eight batches** | 0.5946 (94376/158711) | 0.4239 (67282/158711) | 0.3244 (51488/158711) | 0.2303 (36557/158711) | 0.1718 (27271/158711) | 0.1004 (15931/158711) | 0.0640 (10155/158711) |
| **⑤ El Zahed 663** | 0.5154 (334/648) | 0.3302 (214/648) | 0.2531 (164/648) | 0.1667 (108/648) | 0.1003 (65/648) | 0.0602 (39/648) | 0.0370 (24/648) |
| *Baseline (= 1 − cut-off)* | *0.50* | *0.30* | *0.20* | *0.10* | *0.05* | *0.02* | *0.01* |

## Note S9 · Pairwise AUC for the six library pairs, with shuffled-label baselines

Main-text location: Results. Not on disk; contents taken from the SI Inventory, Section 5.

Table S17 · Pairwise AUC for the six library pairs

| Pair | AUC |
|---|---|
| ②Le Goff × ④ChEMBL | **0.9406** |
| ①CO-ADD × ②Le Goff | **0.9132** |
| ②Le Goff × ③El Zahed | **0.8875** |
| ③El Zahed × ④ChEMBL | **0.8673** |
| ①CO-ADD × ④ChEMBL | **0.8484** |
| ①CO-ADD × ③El Zahed | **0.8092** |

The lowest cell is the random forest for ①CO-ADD × ③El Zahed, 0.7483 [0.7210, 0.7742]; the shuffled-label baselines lie in 0.4631–0.5379.

Table S18 · Three-row comparison for generative models

| Comparison | Cells | AUC |
|---|---|---|

| | | |
|---|---|---|
| Real library × real library | 18 | **0.8558** |
| Generated set × another group's real library | 27 | **0.9033** |
| Generated set × its own training corpus | 12 | **0.7848** |

Three points must be stated: the numbers of cells and the sample sizes differ, so this is not a controlled comparison; none of the 57 cells crosses 0.5; and what is being measured is whether a molecule falls inside the training distribution, not chemical novelty.

Table S19 · Intermediate counts for the 3,661-molecule set

| **Intermediate quantity** | **Value** |
|---|---|
| First | **232** |
| Second | **3,429** |
| Third | **120** |

The denominator is 3,661, not 3,781.

## Note S10 · Single-library baselines across ten libraries

Main-text location: Discussion. Source file: singlelib_baseline_A2_SB2.tsv (md5 83b5f4bd42fd925d8689550c3875b2ac, skiprows=15) and the companion .md (md5 543507dfba15772a9eb705a1db917463).

Table note: the within-library values are taken from the companion .md (the tsv itself is a pure cross-library table); the cross-library median is over 27 cells per library rather than 30, because ④b El Zahed (3,779) has a single class and its AUC is empty.

Further notes

• One model is trained per library and compared on cross-library metrics; the within-library OOF AUC must not be used to select a model — the within-library values are given for reference only.

• CV = RepeatedStratifiedKFold(5 folds × 5 repeats, seed=42); models: logreg(L2,C=1) / rf(depth3,n100) / xgb(depth3,n200,lr0.1); bootstrap nboot=2000, seed=42; shuffled-label baseline SHUF_SEED=20260823.

• The intersection between the training and test libraries is removed first (from the test side). Features = 16 columns (no version of min_proj_area is included).

• Stated limitation: the 3,779 El Zahed molecules (all with wild-type EC50 (μM) of >50) were ruled out as negatives — it cannot be distinguished whether a molecule fails to enter or is simply not an antibacterial. El Zahed is evaluated twice here: ④ mixed (722 positive + 3,779 negative) and ④b, the 3,779 alone (a single class, so no AUC can be determined). As a training library ④b is likewise single-class and cannot be trained; it has been skipped.

• The positive-to-negative ratio differs by a factor of 300 across the six libraries (CO-ADD 0.33% vs the multi-porin measure 55%), so the fraction called correctly and the enrichment are not comparable; only AUC is reported here.

• Rows are ordered by library index ①→⑩; they are not sorted by readout and are not ranked.

Table S20 · Single-library baselines (cross-library AUC)

| Training library | n (pos/neg) | Cells | Median cross-library AUC | Lowest | P25 | Highest | Lowest cell | Median within-library OOF | Gap (pp) |
|---|---|---|---|---|---|---|---|---|---|
| **①CO-ADD 74k** | 72614 (239/72375) | 27 | 0.6023 | 0.3632 | 0.5743 | 0.8646 | ⑤the multi-porin measure 62 / XGBoost | 0.8289 | +22.7 |
| **②Ritz 6,416** | 6388 (2868/3520) | 27 | 0.5062 | 0.1701 | 0.4046 | 0.6254 | ⑤the multi-porin measure 62 / logistic regression | 0.8199 | +31.4 |
| **③ChEMBL *E.coli*** | 16237 (4582/11655) | 27 | 0.5891 | 0.2488 | 0.5638 | 0.8066 | ⑤the multi-porin measure 62 / XGBoost | 0.7024 | +11.3 |
| **④El Zahed (mixed evaluation)** | 4381 (658/3723) | 27 | 0.5909 | 0.3254 | 0.4958 | 0.7743 | ②Ritz 6,416 / XGBoost | 0.6957 | +10.5 |
| **⑤the multi-porin measure 62** | 59 (32/27) | 27 | 0.4810 | 0.3451 | 0.4211 | 0.5941 | ⑦MDAGS Stokes S1B / XGBoost | 0.8009 | +32.0 |
| **⑥Collins DataS2 339** | 339 (53/286) | 27 | 0.5462 | 0.1551 | 0.4745 | 0.6718 | ⑤the multi-porin measure 62 / logistic regression | 0.6996 | +15.3 |
| **⑦MDAGS Stokes S1B** | 2268 (116/2152) | 27 | 0.5916 | 0.3644 | 0.5587 | 0.7702 | ⑤the multi-porin measure 62 / XGBoost | 0.8053 | +21.4 |
| **⑧AMD MIC_Ecoli** | 19875 (491/19384) | 27 | 0.6506 | 0.5253 | 0.6063 | 0.8055 | ②Ritz 6,416 / logistic regression | 0.6705 | +2.0 |
| **⑨eNTRy+Accum** | 361 (96/265) | 27 | 0.5679 | 0.2653 | 0.5071 | 0.7331 | ⑤the multi-porin measure 62 / random forest | 0.9231 | +35.5 |
| **⑩AMD MIC_Saureus** | 32510 (808/31702) | 27 | 0.6421 | 0.5179 | 0.5885 | 0.7520 | ④El Zahed (mixed evaluation) / XGBoost | 0.6632 | +2.1 |

Table S21 · Within-library OOF (three models per library)

| Training library | Model | Within-library OOF AUC | CI | Shuffled baseline |
|---|---|---|---|---|
| **①CO-ADD 74k** | logistic regression(L2, C=1) | 0.7990 | [0.7651, 0.8335] | 0.5528 |
| | random forest(depth=3, n=100, single-threaded) | 0.8289 | [0.7987, 0.8596] | 0.5227 |

| | | | | |
|---|---|---|---|---|
| | XGBoost(depth=3, n=200, lr=0.1, single-threaded) | 0.8881 | [0.8642, 0.9100] | 0.5264 |
| **②Ritz 6,416** | logistic regression(L2, C=1) | 0.8199 | [0.8094, 0.8301] | 0.4974 |
| | random forest(depth=3, n=100, single-threaded) | 0.8058 | [0.7951, 0.8159] | 0.4972 |
| | XGBoost(depth=3, n=200, lr=0.1, single-threaded) | 0.8514 | [0.8422, 0.8603] | 0.4965 |
| **③ChEMBL *E.coli*** | logistic regression(L2, C=1) | 0.6599 | [0.6507, 0.6696] | 0.4846 |
| | random forest(depth=3, n=100, single-threaded) | 0.7024 | [0.6933, 0.7111] | 0.5007 |
| | XGBoost(depth=3, n=200, lr=0.1, single-threaded) | 0.8037 | [0.7961, 0.8111] | 0.4997 |
| **④El Zahed (mixed evaluation)** | logistic regression(L2, C=1) | 0.6933 | [0.6699, 0.7150] | 0.5031 |
| | random forest(depth=3, n=100, single-threaded) | 0.6957 | [0.6719, 0.7191] | 0.4968 |
| | XGBoost(depth=3, n=200, lr=0.1, single-threaded) | 0.7471 | [0.7238, 0.7689] | 0.5131 |
| **⑤the multi-porin measure 62** | logistic regression(L2, C=1) | 0.8009 | [0.6759, 0.9086] | 0.4201 |
| | random forest(depth=3, n=100, single-threaded) | 0.7951 | [0.6620, 0.9062] | 0.4155 |
| | XGBoost(depth=3, n=200, lr=0.1, single-threaded) | 0.8322 | [0.7188, 0.9225] | 0.3530 |
| **⑥Collins DataS2 339** | logistic regression(L2, C=1) | 0.6996 | [0.6236, 0.7741] | 0.5402 |
| | random forest(depth=3, n=100, single-threaded) | 0.6794 | [0.5954, 0.7679] | 0.5201 |
| | XGBoost(depth=3, n=200, lr=0.1, single-threaded) | 0.7543 | [0.6757, 0.8310] | 0.4524 |
| **⑦MDAGS Stokes S1B** | logistic regression(L2, C=1) | 0.7942 | [0.7472, 0.8399] | 0.5190 |
| | random forest(depth=3, n=100, single-threaded) | 0.8053 | [0.7522, 0.8541] | 0.4746 |
| | XGBoost(depth=3, n=200, lr=0.1, single-threaded) | 0.8211 | [0.7630, 0.8712] | 0.5289 |
| **⑧AMD MIC_Ecoli** | logistic regression(L2, C=1) | 0.6578 | [0.6311, 0.6846] | 0.5055 |
| | random forest(depth=3, n=100, single-threaded) | 0.6705 | [0.6452, 0.6965] | 0.5277 |
| | XGBoost(depth=3, n=200, lr=0.1, single-threaded) | 0.7067 | [0.6808, 0.7327] | 0.5253 |

| | | | | |
|---|---|---|---|---|
| **⑨eNTRy+Accum** | logistic regression(L2, C=1) | 0.7934 | [0.7385, 0.8427] | 0.3937 |
| | random forest(depth=3, n=100, single-threaded) | 0.9231 | [0.8899, 0.9500] | 0.4511 |
| | XGBoost(depth=3, n=200, lr=0.1, single-threaded) | 0.9578 | [0.9291, 0.9801] | 0.4341 |
| **⑩AMD MIC_Saureus** | logistic regression(L2, C=1) | 0.6592 | [0.6383, 0.6791] | 0.4884 |
| | random forest(depth=3, n=100, single-threaded) | 0.6632 | [0.6436, 0.6834] | 0.4970 |
| | XGBoost(depth=3, n=200, lr=0.1, single-threaded) | 0.7146 | [0.6956, 0.7341] | 0.5171 |

## Note S11 · The mechanism combination crossed with the activity measure

The three permeation measures are held at their own decision cut-offs (0.95 / 0.98 / 0.99) and combined by union; the efflux percentile runs across the columns and the activity cut-off down the rows. The last row is the activity measure at its native threshold, a score rather than a percentile. Every cell gives precision / n / recall. Recall is taken on 175 positives.

Denominator. All readouts here are on 442 molecules (175 positive / 267 negative, baseline 0.396), obtained by removing from the marketed reference set every molecule that appears in the training set of any of the five measures. This subset is not the one used in the main text (515), and the two are not to be read against each other.

Purpose. The table is a sensitivity analysis: it is reported in full so that the range over which the statements in the Discussion hold can be read off. No cell is selected, the rows are not sorted, and no cut-off is recommended.

Table S22 · Molecules called by both the mechanism combination and the activity measure

| **Activity cut-off** | **0.50** | **0.70** | **0.80** | **0.90** | **0.95** | **0.98** | **0.99** |
|---|---|---|---|---|---|---|---|
| **0.50** | 0.796 / 103 / 0.469 | 0.796 / 103 / 0.469 | 0.796 / 103 / 0.469 | 0.812 / 101 / 0.469 | 0.845 / 97 / 0.469 | 0.907 / 86 / 0.446 | 0.907 / 75 / 0.389 |
| **0.70** | 0.818 / 99 / 0.463 | 0.818 / 99 / 0.463 | 0.818 / 99 / 0.463 | 0.835 / 97 / 0.463 | 0.853 / 95 / 0.463 | 0.906 / 85 / 0.440 | 0.905 / 74 / 0.383 |
| **0.80** | 0.825 / 97 / 0.457 | 0.825 / 97 / 0.457 | 0.825 / 97 / 0.457 | 0.842 / 95 / 0.457 | 0.860 / 93 / 0.457 | 0.917 / 84 / 0.440 | 0.918 / 73 / 0.383 |
| **0.90** | 0.840 / 94 / 0.451 | 0.840 / 94 / 0.451 | 0.840 / 94 / 0.451 | 0.859 / 92 / 0.451 | 0.868 / 91 / 0.451 | 0.916 / 83 / 0.434 | 0.917 / 72 / 0.377 |
| **0.95** | 0.852 / 88 / 0.429 | 0.852 / 88 / 0.429 | 0.852 / 88 / 0.429 | 0.872 / 86 / 0.429 | 0.882 / 85 / 0.429 | 0.935 / 77 / 0.411 | 0.925 / 67 / 0.354 |
| **0.98** | 0.944 / 71 / 0.383 | 0.944 / 71 / 0.383 | 0.944 / 71 / 0.383 | 0.944 / 71 / 0.383 | 0.957 / 70 / 0.383 | 0.956 / 68 / 0.371 | 0.948 / 58 / 0.314 |

| | | | | | | | |
|---|---|---|---|---|---|---|---|
| **0.99** | 0.984 / 63 / 0.354 | 0.984 / 63 / 0.354 | 0.984 / 63 / 0.354 | 0.984 / 63 / 0.354 | 0.984 / 63 / 0.354 | 0.984 / 62 / 0.349 | 0.981 / 52 / 0.291 |
| **score ≥ 0.5** | 0.984 / 63 / 0.354 | 0.984 / 63 / 0.354 | 0.984 / 63 / 0.354 | 0.984 / 63 / 0.354 | 0.984 / 63 / 0.354 | 0.984 / 62 / 0.349 | 0.981 / 52 / 0.291 |

Table note: cell = precision / n / recall.

Table S23 · Molecules called by the mechanism combination only

| Activity cut-off | 0.50 | 0.70 | 0.80 | 0.90 | 0.95 | 0.98 | 0.99 |
|---|---|---|---|---|---|---|---|
| **0.50** | 0.583 / 24 / 0.080 | 0.471 / 17 / 0.046 | 0.333 / 12 / 0.023 | 0.364 / 11 / 0.023 | 0.400 / 10 / 0.023 | 0.375 / 8 / 0.017 | 0.500 / 2 / 0.006 |
| **0.70** | 0.536 / 28 / 0.086 | 0.429 / 21 / 0.051 | 0.312 / 16 / 0.029 | 0.333 / 15 / 0.029 | 0.417 / 12 / 0.029 | 0.444 / 9 / 0.023 | 0.667 / 3 / 0.011 |
| **0.80** | 0.533 / 30 / 0.091 | 0.435 / 23 / 0.057 | 0.333 / 18 / 0.034 | 0.353 / 17 / 0.034 | 0.429 / 14 / 0.034 | 0.400 / 10 / 0.023 | 0.500 / 4 / 0.011 |
| **0.90** | 0.515 / 33 / 0.097 | 0.423 / 26 / 0.063 | 0.333 / 21 / 0.040 | 0.350 / 20 / 0.040 | 0.438 / 16 / 0.040 | 0.455 / 11 / 0.029 | 0.600 / 5 / 0.017 |
| **0.95** | 0.538 / 39 / 0.120 | 0.469 / 32 / 0.086 | 0.407 / 27 / 0.063 | 0.423 / 26 / 0.063 | 0.500 / 22 / 0.063 | 0.529 / 17 / 0.051 | 0.700 / 10 / 0.040 |
| **0.98** | 0.518 / 56 / 0.166 | 0.469 / 49 / 0.131 | 0.432 / 44 / 0.109 | 0.463 / 41 / 0.109 | 0.514 / 37 / 0.109 | 0.615 / 26 / 0.091 | 0.737 / 19 / 0.080 |
| **0.99** | 0.531 / 64 / 0.194 | 0.491 / 57 / 0.160 | 0.462 / 52 / 0.137 | 0.490 / 49 / 0.137 | 0.545 / 44 / 0.137 | 0.625 / 32 / 0.114 | 0.720 / 25 / 0.103 |
| **score ≥ 0.5** | 0.531 / 64 / 0.194 | 0.491 / 57 / 0.160 | 0.462 / 52 / 0.137 | 0.490 / 49 / 0.137 | 0.545 / 44 / 0.137 | 0.625 / 32 / 0.114 | 0.720 / 25 / 0.103 |

Table note: cell = precision / n / recall. Cells with fewer than 15 positives are the smallest in this table; the smallest cell holds 2 molecules, and readouts from this set are not to be quoted without n.

Table S24 · Molecules called by the activity measure only

| Activity cut-off | 0.50 | 0.70 | 0.80 | 0.90 | 0.95 | 0.98 | 0.99 |
|---|---|---|---|---|---|---|---|
| **0.50** | 0.267 / 217 / 0.331 | 0.267 / 217 / 0.331 | 0.267 / 217 / 0.331 | 0.265 / 219 / 0.331 | 0.260 / 223 / 0.331 | 0.265 / 234 / 0.354 | 0.294 / 245 / 0.411 |
| **0.70** | 0.304 / 191 / 0.331 | 0.304 / 191 / 0.331 | 0.304 / 191 / 0.331 | 0.301 / 193 / 0.331 | 0.297 / 195 / 0.331 | 0.302 / 205 / 0.354 | 0.333 / 216 / 0.411 |
| **0.80** | 0.329 / 164 / 0.309 | 0.329 / 164 / 0.309 | 0.329 / 164 / 0.309 | 0.325 / 166 / 0.309 | 0.321 / 168 / 0.309 | 0.322 / 177 / 0.326 | 0.356 / 188 / 0.383 |
| **0.90** | 0.395 / 129 / 0.291 | 0.395 / 129 / 0.291 | 0.395 / 129 / 0.291 | 0.389 / 131 / 0.291 | 0.386 / 132 / 0.291 | 0.386 / 140 / 0.309 | 0.424 / 151 / 0.366 |
| **0.95** | 0.484 / 93 / 0.257 | 0.484 / 93 / 0.257 | 0.484 / 93 / 0.257 | 0.474 / 95 / 0.257 | 0.469 / 96 / 0.257 | 0.462 / 104 / 0.274 | 0.509 / 114 / 0.331 |
| **0.98** | 0.678 / 59 / 0.229 | 0.678 / 59 / 0.229 | 0.678 / 59 / 0.229 | 0.678 / 59 / 0.229 | 0.667 / 60 / 0.229 | 0.677 / 62 / 0.240 | 0.722 / 72 / 0.297 |
| **0.99** | 0.681 / 47 / 0.183 | 0.681 / 47 / 0.183 | 0.681 / 47 / 0.183 | 0.681 / 47 / 0.183 | 0.681 / 47 / 0.183 | 0.688 / 48 / 0.189 | 0.741 / 58 / 0.246 |

| | | | | | | | |
|---|---|---|---|---|---|---|---|
| **score ≥ 0.5** | 0.706 / 51 / 0.206 | 0.706 / 51 / 0.206 | 0.706 / 51 / 0.206 | 0.706 / 51 / 0.206 | 0.706 / 51 / 0.206 | 0.712 / 52 / 0.211 | 0.758 / 62 / 0.269 |

Table note: cell = precision / n / recall.

Table S25 · The mechanism combination on its own

| Efflux cut-off | n | precision | recall |
|---|---|---|---|
| **0.50** | 127 | 0.756 | 0.549 |
| **0.70** | 120 | 0.750 | 0.514 |
| **0.80** | 115 | 0.748 | 0.491 |
| **0.90** | 112 | 0.768 | 0.491 |
| **0.95** | 107 | 0.804 | 0.491 |
| **0.98** | 94 | 0.862 | 0.463 |
| **0.99** | 77 | 0.896 | 0.394 |

Table note: the mechanism combination does not depend on the activity cut-off, so its seven rows are the same in every row of Tables S21 to S23.

Table S26 · The activity measure on its own

| Activity cut-off | n | precision | recall |
|---|---|---|---|
| **0.50** | 320 | 0.438 | 0.800 |
| **0.70** | 290 | 0.479 | 0.794 |
| **0.80** | 261 | 0.513 | 0.766 |
| **0.90** | 223 | 0.583 | 0.743 |
| **0.95** | 181 | 0.663 | 0.686 |
| **0.98** | 130 | 0.823 | 0.611 |
| **0.99** | 110 | 0.855 | 0.537 |
| **score ≥ 0.5** | 114 | 0.860 | 0.560 |

Table note: the activity measure does not depend on the efflux cut-off, so its eight rows are the same in every column of Tables S21 to S23.

Paired bootstrap. Differences in precision were resampled on the object set only, with the models left untrained (B = 2,000, seed = 42; no missing difference and no empty positive call in any resample). The difference between the both-called set and the mechanism combination excludes zero in 51 of the 56 cells; the five that include zero all lie at efflux 0.99 with the activity measure at the 0.95 percentile or looser, where the mechanism side has already been taken to its tightest and the both-called set is close to the mechanism combination itself. The difference between the both-called set and the activity measure excludes zero in all 56 cells. The difference between the activity measure and the mechanism combination includes zero in 15 cells, and among the remaining 41 the sign is not constant: the mechanism

combination is the more precise of the two where the activity measure is loose, and the activity measure is the more precise where it is tight and the efflux cut-off is loose.

## Note S12 · Per-measure recall and denominator on the El Zahed 650 set

Main-text location: Data collection. Not on disk; contents taken from SI Inventory, Section 5.

Table S27 · Per-measure recall on the El Zahed 650 set

| Measure | Recall | Denominator |
|---|---|---|
| the OmpF measure | 14.0% | 662 |
| the efflux measure | 15.0% | 660 |
| the activity measure | 6.2% | 650 |
| the lipid route measure | 3.7% | 650 |
| the multi-porin measure | 3.4% | 650 |



\end{document}